\documentclass[10pt, conference, compsocconf]{IEEEtran}
\usepackage{fixme}
\usepackage{cite}
\usepackage{graphicx}
\usepackage{amssymb}
\usepackage{adjustbox}
\ifCLASSINFOpdf
\else
\fi
\usepackage[cmex10]{amsmath}
\usepackage{algorithmic}

\usepackage{array}

\usepackage{mdwmath}
\usepackage{mdwtab}

\usepackage[tight,footnotesize]{subfigure}

\usepackage[font=footnotesize]{subfig}
\usepackage{url}

\usepackage[normalem]{ulem}
\useunder{\uline}{\ul}{}
\usepackage{multirow}

\newcolumntype{H}{>{\centering\arraybackslash}m{5.2cm}}
\newcolumntype{Y}{>{\raggedright\arraybackslash}m{5.2cm}}
\newcolumntype{J}{>{\centering\arraybackslash}m{6.7cm}}
\newcolumntype{K}{>{\raggedright\arraybackslash}m{6.8cm}}

\renewcommand{\raggedright}{\leftskip=2pt \rightskip=2pt}
\newcolumntype{C}{>{\centering\arraybackslash}m{2.5cm}}
\newcolumntype{D}{>{\centering\arraybackslash}m{4.5cm}}
\newcolumntype{L}{>{\raggedright\arraybackslash}m{4cm}}
\newcolumntype{T}{>{\raggedright\arraybackslash}m{4.5cm}}
\newcolumntype{W}{>{\raggedright\arraybackslash}m{14cm}}

\begin{document}
%
\title{ Solving Cold-Start Problem in Large-scale Recommendation Engines: \\ A Deep Learning Approach}


\author{\IEEEauthorblockN{Jianbo Yuan\IEEEauthorrefmark{1},
		Walid Shalaby\IEEEauthorrefmark{4},
		Mohammed Korayem\IEEEauthorrefmark{3},
		David Lin\IEEEauthorrefmark{3},
		Khalifeh AlJadda\IEEEauthorrefmark{3},
		and Jiebo Luo \IEEEauthorrefmark{1}
		}
	\IEEEauthorblockA{\IEEEauthorrefmark{1}Department of Computer Science, 
		University of Rochester,
		New York\\ Email: jyuan10, jluo@cs.rochester.edu}
	\IEEEauthorblockA{\IEEEauthorrefmark{4} Department of Computer Science,  
		University of North Carolina at Charlotte\\
		Email: wshalaby@uncc.edu}
	\IEEEauthorblockA{\IEEEauthorrefmark{3}CareerBuilder, Norcross,GA\\
		Email: mohammed.korayem, david.lin, khalifeh.aljadda@careerbuilder.com}
	
	}


%


\newcommand{\mustfix}[1]{\fixme{\hl{#1}}}
\newcommand{\pleasenote}[1]{\fxnote{\hl{#1}}}
\newcommand{\hlfixme}[1]{\fixme{\hl{#1}}}
\newcommand{\hlfxnote}[1]{\fxnote{\hl{#1}}}

\maketitle

\begin{abstract}
Collaborative Filtering (CF) is widely used in large-scale recommendation engines because of its efficiency, accuracy and scalability. However, in practice, the fact that recommendation engines based on CF require interactions between users and items before making recommendations, make it inappropriate for new items which haven't been exposed to the end users to interact with. This is known as the cold-start problem.
In this paper we introduce a novel approach which employs deep learning to tackle this problem in any CF based recommendation engine. One of the most important features of the proposed technique is the fact that it can be applied on top of any existing CF based recommendation engine without changing the CF core. We successfully applied this technique to overcome the \emph{item} cold-start problem in Careerbuilder's CF based recommendation engine. Our experiments show that the proposed technique is very efficient to resolve the cold-start problem while maintaining high accuracy of the CF recommendations.
\end{abstract}


%
\IEEEpeerreviewmaketitle

\section{Introduction}

\label{intro}
Recommendation Systems (RSs)  utilize knowledge discovery and data mining techniques in order to predict items of interest to individual users and subsequently suggest these items to them as recommendations \cite{bobadilla2013recommender,lu2015recommender}. Over the years, techniques and applications of RSs have evolved in both academia and industry due to the exponential increase in the number of both on-line information and on-line users. Such volumes of big data generated at very high velocities pose many challenges when it comes to developing and deploying scalable accurate RSs. Applications domains of RSs include: ``e-government, e-business, e-commerce/e-shopping, e-library, e-learning, e-tourism, e-resource services and e-group activities" \cite{lu2015recommender}. On the other hand, new RSs deployment platforms emerged over years to include, besides the classical Web-based platform, other modern platforms like mobile, TV, and radio \cite{lu2015recommender}. Depending on the application domain and the deployment platform, dozens of techniques and methods have been proposed to address cross-domain and cross-platform scalability and quality challenges \cite{bobadilla2013recommender,lops2011content,su2009survey}.

Approaches for building RSs can be broadly clustered into three main categories: Content-Based (CB), Collaborative Filtering (CF), and hybrid techniques. CB techniques work by measuring or predicting the similarity between profiles of the items (attributes/descriptions) and profiles of the users' (attributes/descriptions of past preferred items) \cite{aggarwal2016content,pazzani2007content}. Typically, the similarity is either measured using a suitable function (e.g., cosine similarity), or predicted using statistical or machine learning models (e.g., regression). CB systems are therefore suitable in domains where items' descriptions or their metadata can be easily obtained (e.g., books, news articles, jobs, TV shows, etc.). They are also suitable when the lifetime of items is short and/or the change in the item recommendation pool is frequent due to many new items entering the pool over short periods. These constraints limit the number of ratings those items can receive over their lifetime. Many challenges arise when it comes to CB systems. For example, items' metadata might be very limited or does not really contribute to generating user's interest. Item description, on the other hand, is typically textual which makes the similarity scoring more challenging due to language ambiguity raising the need for semantic-aware CB systems \cite{de2015semantics}.

CF is perhaps the most prominent and successful techniques of modern recommendation systems. Unlike CB methods which recommend items that are similar to what target user liked in the past, CF methods leverage preferences of other similar users in order to make recommendations to the target user \cite{aggarwal2016content,su2009survey}. In other words, CF tries to predict good recommendations by analyzing the correlations between all user behaviors rather than analyzing correlations between items content. CF is therefore suitable in domains where obtaining meta-features or descriptions of items is infeasible. Another motivation for CF is to introduce more diversity and serendipity to user's experience by leveraging the wisdom of the crowd, i.e., recommending items that the target user might not have thought about or consumed before but rather liked by other similar users. One of the major challenges of CF based recommendation systems is the cold-start problem which occurs when there is a lack of information linking new users or items. In such cases, the RS is unable to determine how those users or items are related and is therefore unable to provide useful recommendations. Hybrid RSs try to achieve better performance by combining two or more techniques. The main theme of hybrid RSs is to leverage the advantages of both CB and CF, and to alleviate their shortcomings at the same time through hybridization.

\begin{figure}[th]
	\centering
	\includegraphics[width=3.2in]{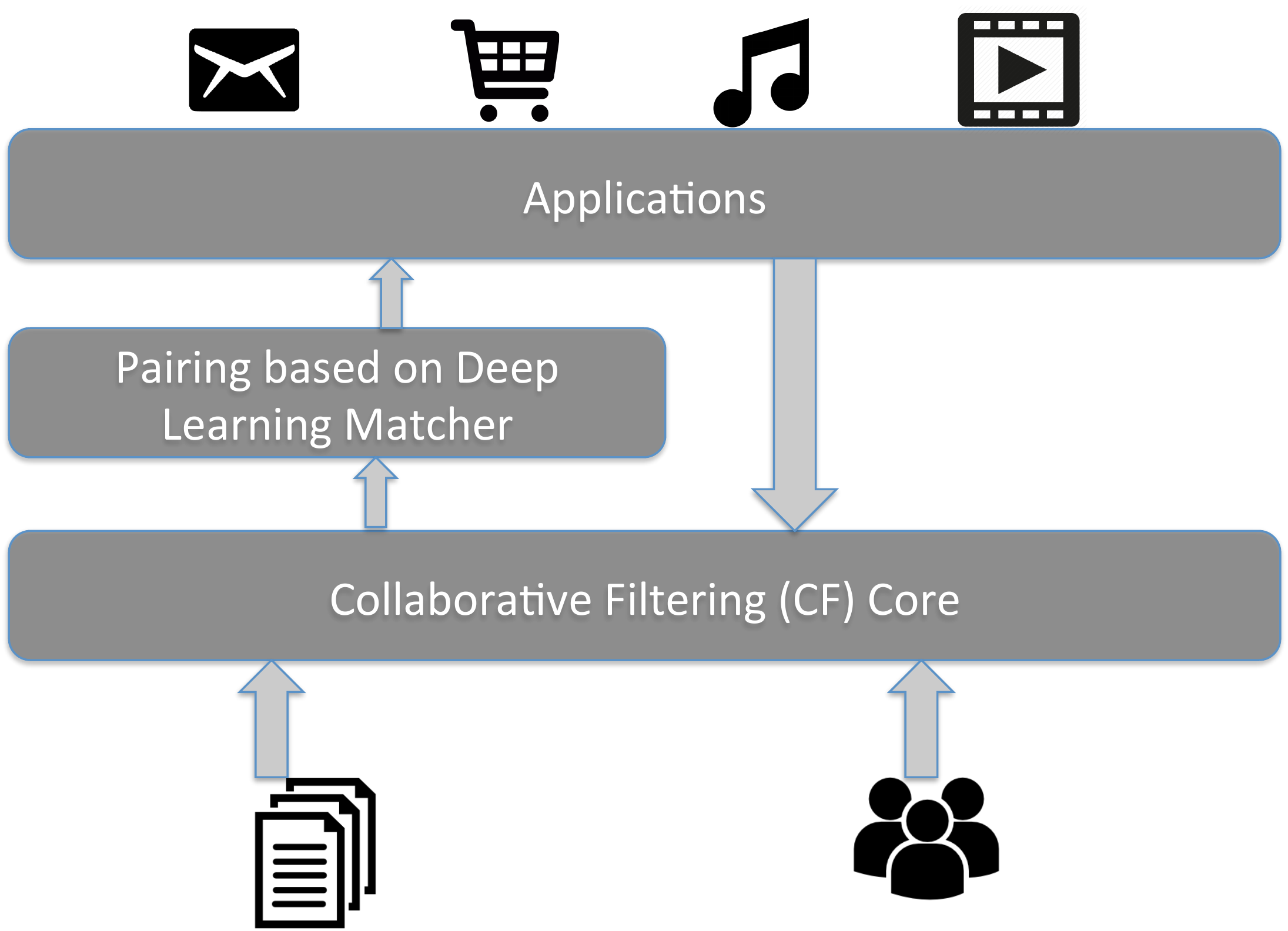}	
	\caption{System architecture of CF based recommendation engine with the proposed pairing layer to solve the cold-start problem. An application from the application's layer sends a recommendation request to the CF core which analyzes the interactions between users and items to generate recommendations. Once those recommendations are generated they are sent to the pairing layer. The pairing layer will check the list of recommended items and if any item among the recommendation list has been paired with a new item, then the new item will be added to the recommendation list which is passed to the application layer. }
	\label{layers}
\end{figure}

In this work, we propose a deep learning based solution to the cold-start problem. Our solution utilizes a state-of-the-art deep learning document embedding algorithms (also known as \emph{doc2vec}) \cite{le2014distributed}. Our main contributions are as follows:
\begin{itemize}
	\item We propose a deep learning based matching algorithm to solve cold-start and sparsity problems in CF based recommendation systems without major changes in the existing system as shown in figure \ref{layers}.
	\item We improve the performance of the deep learning matcher by incorporating contextual meta-data which boost the accuracy significantly.
	\item The proposed algorithm is capable of being extended to solve similar cold-start problems in different real application scenarios such as relevancy and ranking in search engines.
	
\end{itemize}

The remainder of the paper is organized as follows: in Section \ref{related} we will introduce the related work in recommendation systems and the document embedding; Section \ref{background} concludes CF related techniques; we discuss the problem description and proposed algorithm in Section \ref{method}; Section \ref{exp} will explain the experiments, case study, and discussion. Finally we conclude in Section \ref{discussion}.

\section{Related Work}
\label{related}

The main objective for the Recommendation Systems (RSs) is providing a user with content he/she would like by estimating the relevancy or the rating of these contents based on the information about users and the items \cite{bobadilla2013recommender,adomavicius2005toward}.
The cold-start problem is one of the major challenges in RSs design and deployment. Cold-start occurs when either new users or new items are introduced into the system. In this situation there would be no behavioral data (ratings, purchases, clicks, etc.) for CF to work properly. Addressing cold-start is inevitable in modern RSs for two reasons \cite{saveski2014item}: 1) the item and user pools change on daily basis, and 2) CF is considered state-of-the-art recommendation technique but it requires significant behavioral data in order to work properly. Therefore, it is important to promote new items as quickly as possible in order to establish links between them and users to improve CF performance subsequently.

Another related problem to CF is the data sparsity which appears due to insufficient ratings per user and/or item in the rating matrix challenging the ability of CF to predict accurate user preferences and item similarities.

Several methods have been introduced to address the cold-start and the data sparsity problems \cite{schein2002methods,guo2013integrating,bobadilla2012collaborative,saveski2014item,soboroff1999combining}. In this work, we focus on the problem of cold-start in item-based CF. Most of the proposed methods to address item cold-start adopt content-based approach; they utilize the content of new items in order to identify similar user profiles and subsequently recommend these new items to them. Saveski and Mantrach \cite{saveski2014item} proposed Local Collective Embeddings (LCE) as a solution for item cold-start. LCE utilizes descriptions of the new items (i.e., the term-document matrix) and project it collectively with the user-item rating matrix into a common low-dimensional space using non-negative matrix factorization. Soboroff and Nicholas \cite{soboroff1999combining} utilized Latent Semantic Indexing (LSI) \cite{deerwester1990indexing} in order to create topical representations of user profiles in the latent space. New items are then projected into the same latent space and compared to user profiles then recommended to the most similar profiles. Similar to \cite{soboroff1999combining}, Schein et al. \cite{schein2002methods} proposed an approach that creates a joint distribution of items and users through an aspect model in the latent space by combining items' content and users' preferences. Our approach to solve cold-start problem relies on pairing the new items with existing items that have been exposed to the users and gained enough ratings to be considered by CF. Thus, our approach differs from prior ones which try to utilize content-based techniques to pair the new items with users by matching the content of these items with the content of users' profile.


\section{Background}
\label{background}
Collaborative Filtering is one of the
most successful techniques to building RSs due to
their independence from the content of items
being recommended, which make them easy to create
and use across many application domains
\cite{su2009survey}.

Typically, CF methods utilize the user-item
rating matrix $R \in \mathbb{R}^{|U|\times|I|}$
which contains past ratings of items made by
users. Rows in $R$ represent users
$U=\{u_1,u_2,...,u_{|U|}\}$, while columns
represent items $I=\{i_1,i_2,...,i_{|I|}\}$. Each
entry $r_{ui}$ in $R$ represents the rating user
$u$ gave to item $i$. The role of the CF
algorithm is to perform matrix completion filling
empty entries in $R$ by analyzing existing
entries.

The early generation of CF algorithms are the
memory-based (a.k.a neighborhood-based)
techniques
\cite{resnick1994grouplens,konstan1997grouplens}.
These techniques work by measuring the
similarities or correlations between either users
(rows) or items (columns) in the user-item rating
matrix $R$. After finding these similar
neighbors, recommendations are generated by
choosing the top-$K$ items similar to a given
item in case of item-based recommendations, or by
aggregating the correlation scores of items liked
by similar users in case of user-based
recommendations
\cite{sarwar2001item,deshpande2004item,
	linden2003amazon}.

The choice of the similarity or correlation
metric has a major contribution to the quality of
recommendations \cite{sarwar2001item}. {\it
	Pearson-r correlation} is a very popular metric
used in CF systems, and is used to estimate how
well two variables are linearly related. {\it
	Pearson-r correlation $corr_{i,j}$} between two
items or two users $i,j$ is computed as:
\begin{equation} corr_{i,j} = \frac{\sum_{w \in
		W}{(r_{wi}-\bar{r_{i}})(r_{wj}-\bar{r_{j}})}}
{\sqrt{\sum_{w \in
			W}{(r_{wi}-\bar{r_{i}})^2}}{\sqrt{\sum_{w \in
				W}{(r_{wj}-\bar{r_{j}})^2}}}} \end{equation}

where, in case of user-based recommendation,
$w \in W$ denotes items rated by both users
$i,j$. $r_{wi}$ \& $r_{wj}$ are ratings of item
$w$ by users $i,j$ respectively.
$\bar{r_i},\bar{r_j}$ are average ratings of
users $i,j$ respectively.  In case of item-based
recommendation, $w \in W$ denotes users who rated
items $i,j$. $r_{wi}$ \& $r_{wj}$ are ratings of
user $w$ for items $i,j$ respectively.
$\bar{r_i},\bar{r_j}$ are average ratings of
items $i,j$ respectively. Another commonly used
similarity metric is the cosine similarity which
is computed as: \begin{equation} cos_{i,j} =
\frac{\vec{R_i}\ .\ \vec{R_j}}{||\vec{R_i}||\ ||\vec{R_j}||}
\end{equation}

where $\vec{R_i}$ \& $\vec{R_j}$ are the rating vectors
of items/users $i,j$ in the rating matrix $R$ in
case of item-based/user-based recommendation
respectively.

Another category of CF techniques are the
model-based approaches which are more scalable
than memory-based techniques as they offline
build a model of item-item or user-user
similarities and then use it at real-time to
generate recommendations. Several algorithms were
proposed to generate such models including
Bayesian networks
\cite{breese1998empirical,
	miyahara2000collaborative, su2006collaborative},
clustering
\cite{ungar1998clustering,chee2001rectree},
latent semantic analysis
\cite{hofmann2004latent}, Singular Value
Decomposition (SVD) \cite{billsus1998learning},
Alternating Least Squares (ALS)
\cite{zhou2008large,koren2009matrix} regression
models \cite{vucetic2005collaborative}, Markov
Decision Processes (MDPs) \cite{shani2005mdp},
and others \cite{aggarwal2016model}.

Another important aspect to understand this work is how to represent documents in the vector space to measure similarity \cite{turian2010word}. Lexical features are commonly used for extracting vectors from documents including \emph{Bag-of-Words} (\emph{BoW}), \emph{n-grams} (typically bigram and trigram), and term frequency-inverse document frequency (\emph{tf-idf}).
Topic models such as Latent Dirichlet Allocation
(\emph{LDA}) are also used as features in document
classification problems such as sentiment analysis
and have shown promising results
\cite{blei2003latent,maas2011learning}. Moreover, the
application of deep learning to natural language
processing field has shown a great success.

\textbf{\emph{Term Frequency-Inverse Document
		Frequency}}: Common lexical features for document
embedding include \emph{Bag-of-Words}, \emph{n-gram} and
\emph{tf-dif}. Both \emph{Bag-of-Words} and \emph{n-gram}
model draw much attention on frequent words, which may
not be the best way to measure the importance of a word
in a document. On the other hand, \emph{tf-idf} can be
considered as a weighted form of \emph{BoW} for
evaluating the importance of a word to the document. Let
$tf(w;d)$ be the number of times word $w$ appears in
document $d$ from a collection $D$, $idf(w;D)$ indicates
inverse document frequency of word $w$ in set $D$, then
\emph{tf-idf} is defined in Equation \ref{tf} and 4.

\begin{eqnarray}\label{tf} tf-idf&=&tf(w,d) \times
	idf(w,D)\\ idf(w,D)&=&\log\frac{N}{1+\left | \left \{ d\in
		D:w\in d \right \} \right |} \end{eqnarray}

\textbf{\emph{Latent Dirichlet Allocation}}: \emph{LDA}
is a probabilistic model which learns $P(z|w)$, distribution
of a latent topic variable $z$ given a word $w$
\cite{blei2003latent}. Compared with lexical features
(\emph{BoW} and \emph{tf-idf}) mentioned above,
representations learned by \emph{LDA} focus more on the
semantic meanings of each word, and have a feature space that
is in low dimensions. Topic vectors learned represent the
weights of words for each topic, and after normalizing each
word vector from a sentence or a document, we obtain the
vector of the sentence or document for all topics and thus
embed the target document into a vector representation based
on \emph{LDA} model.

\begin{figure*}[t]
	\begin{minipage}[*]{0.5\linewidth}
		\centering
		\includegraphics[height=2.2in]{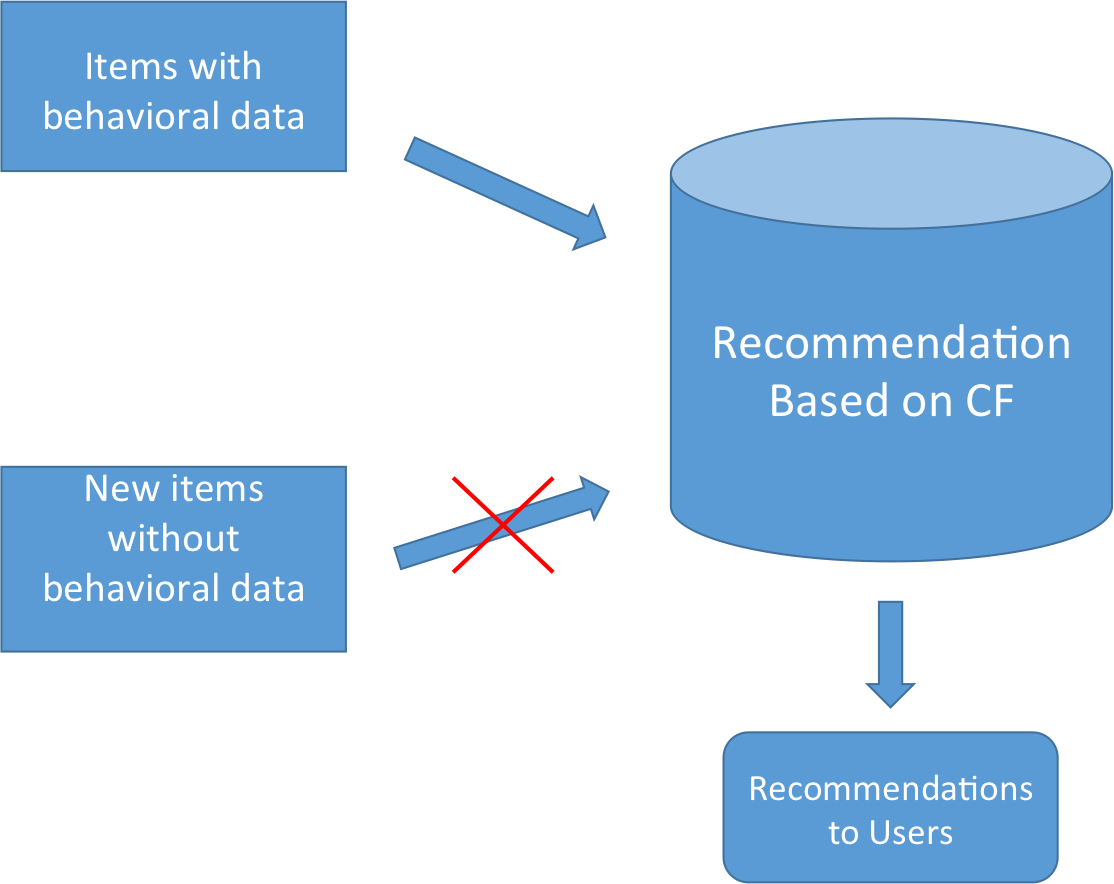}
	\end{minipage}%
	\begin{minipage}[*]{0.5\linewidth}
		\centering
		\includegraphics[height=2.2in]{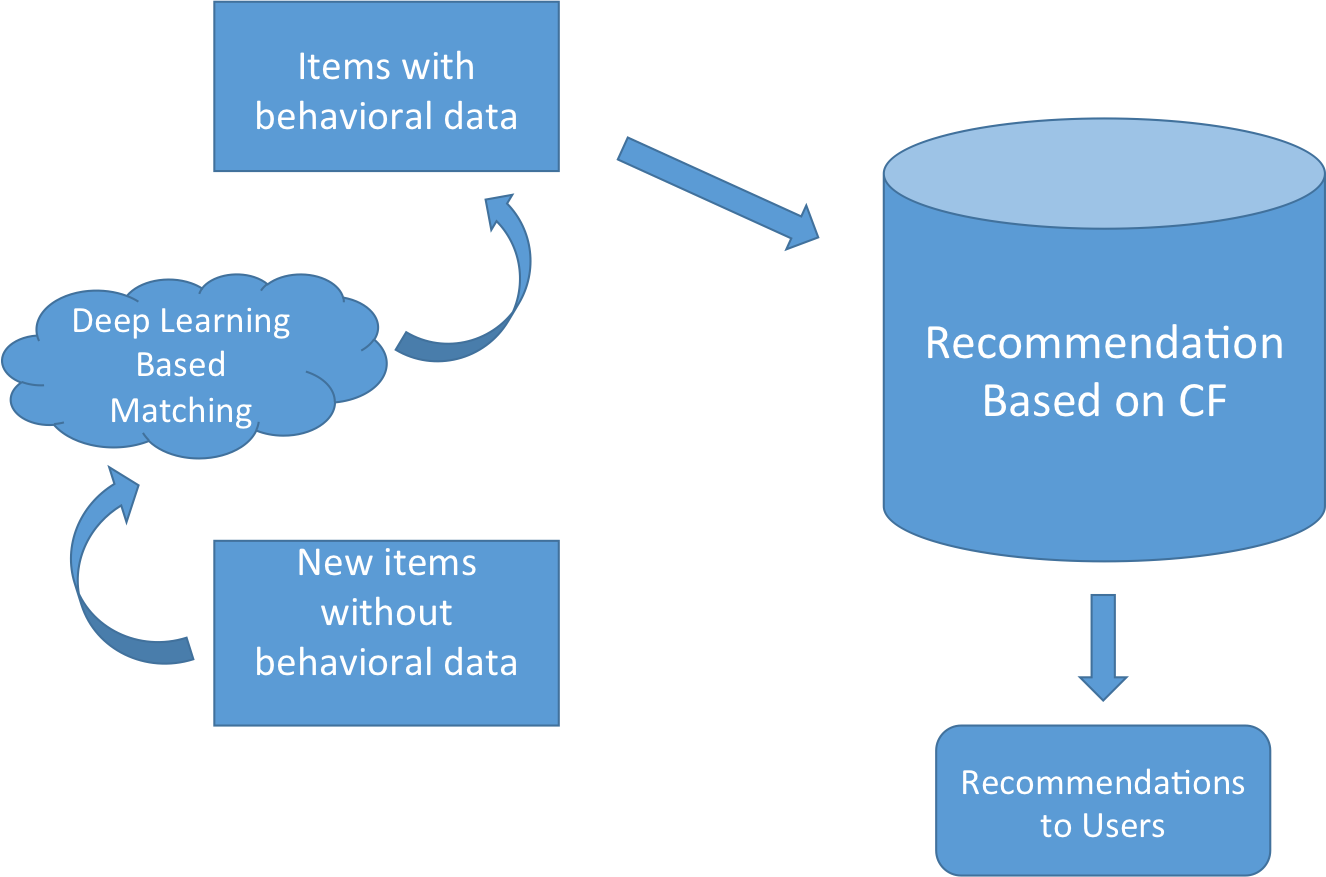}
	\end{minipage}
	\caption{(a) Framework of CF Based Recommendation System: The only items to be considered are those which gained some behavioral data like ratings from users while all the new items are not considered until they gain such behavioral data. (b) Framework of Proposed Pairing Algorithm to Solve the Cold Start Problem: Each new item without behavioral data is paired with existing item(s) that has behavioral data. Once the recommendations are retrieved by CF, each item with behavioral data will pull its pair new item into the recommendation list.}
	\label{frame1}
\end{figure*}

\section{Methodology}
\label{method}

\subsection{Problem Description}
\label{data}
In dynamic domains like job boards, new items appear frequently which makes CF not feasible due to the cold-start problem. However, due to many shortcomings of the CB recommendations, most domains still prefer CF over CB  due to the accuracy and diversity of results CF can provide which CB can not. That said, the cold-start problem continue to impact substantial number of new items. Recommendation engine is considered one of the major channels of engaging users with items and continuing to re-engage them when they leave a website via recommendation emails. Therefore, items which have no chance to be recommended due to the cold-start problem, will lose that major channel of exposure to end users. Thus, we propose a novel technique that helps CF to be considered in dynamic domains with frequent presence of new items by leveraging a deep learning matcher (DLM). The DLM is used to pair each new item with an existing item that has behavioral data (like ratings). This pairing will allow the new items to be considered in CF even before any users interact with these new items.

\subsection{Proposed Framework}
In conventional CF based recommendation engines (Figure \ref{frame1}-a) the only items to be considered for recommendations are those with behavioral data (i.e. users interact with those items by rating, purchasing, clicking...etc), however, all new items without such behavioral data can not be part of the recommendations generated by CF. Our proposed system which is depicted in Figure \ref{frame1}-b adds a new module which can be thought of as a plug-in that will match each new item $i_x$ with an item that has behavioral data $i_y$, we call this process pairing. Once this pairing is done, each pair $(i_x,i_y)$ will be considered as one item, so when item $i_y$ is selected for recommendation by CF, item $i_x$ (the new item with no behavioral data) will appear with that recommendation as well. Therefore, the accuracy of the pairing is very important since pairing irrelevant items will introduce noise to the recommendations.

\begin{figure}[h]
	\centering
	\includegraphics[width=3.2in]{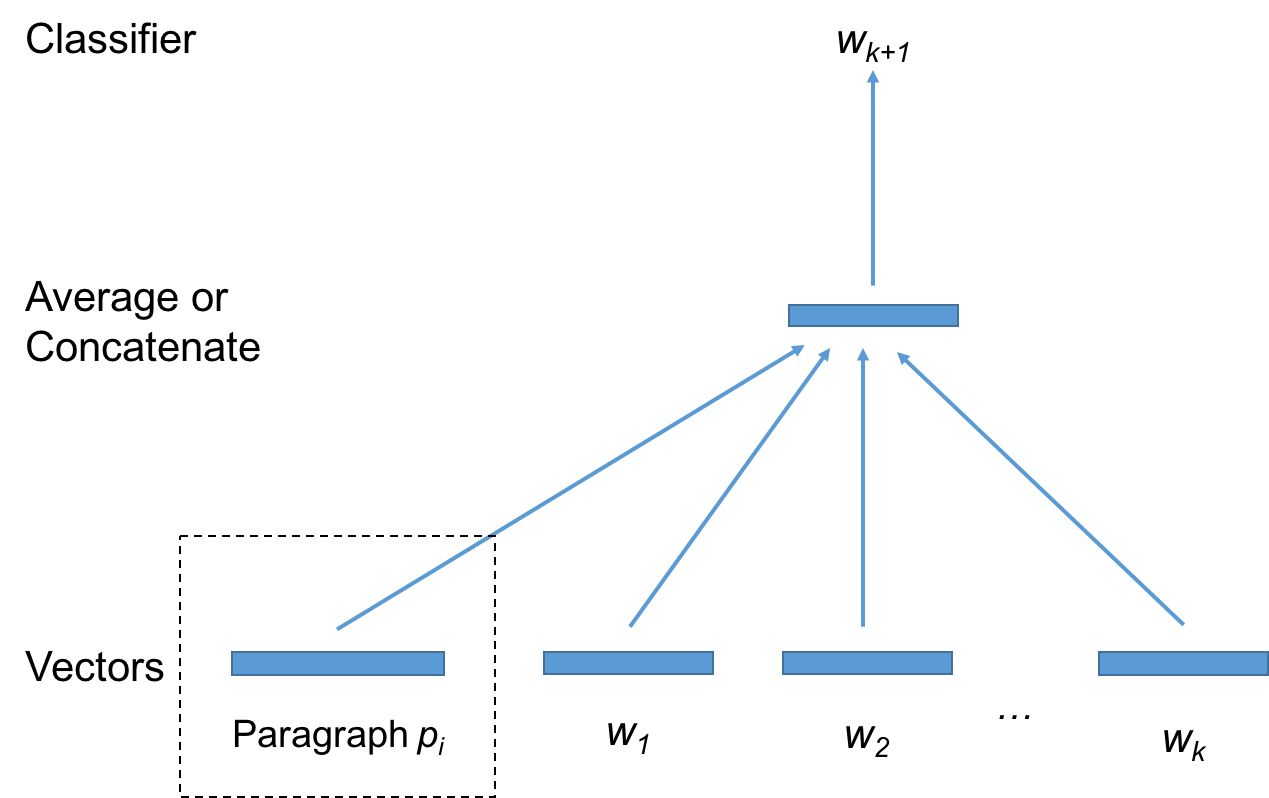}	
	\caption{Learning Word Vector and Document Vector.}
	\label{pg}
\end{figure}

\subsection{Document Embedding and Matching}
Given the importance of accurate pairs (similar ones) for the overall quality of the recommendations to be generated by CF afterwards, we tried the standard document similarity techniques like \emph{Term Frequency-Inverse Document Frequency} (\emph{tf-idf}) and \emph{Latent Dirichlet Allocation} (\emph{LDA}) but the results were not promising. Therefore we built a Deep Learning Matcher (DLM) which outperforms all the other techniques significantly. Our DLM was built utilizing \textbf{\emph{doc2vec}} which is considered the current state-of-the-art deep learning algorithm for document embedding and matching. Contrary to lexical features, the information extracted from each word is distributed all along a word window in distributed representations (as known as \emph{word2vec} and \emph{doc2vec} feature) as shown in Figure \ref{pg} \cite{le2014distributed}. For a word vector learning, given a sequence of $T$ words $\left \{w_{1},w_{2},...,w_{T}\right \}$ and a window size $c$, the objective function is as follows:
\begin{equation}
	\label{obj}
	\frac{1}{T}\sum_{i=c}^{T-c}\log p(w_{i}|w_{i-c},...,w_{i+c})
\end{equation}

In order to maximize the objective function in Equation \ref{obj}, the probability of $w_{i}$ is calculated based on the softmax function shown in Equation \ref{softmax} where the word vectors are concatenated or averaged for predicting the next word in the content. Similarly to learning the word vectors, the processing of learning the paragraph vector (i.e. the \emph{doc2vec} model) is maximizing the averaged log probability with the softmax function by combining the word vectors with the paragraph vector $p_{i}$ in a concatenated or averaged fashion as shown in Figure \ref{pg}. For new documents input in the model, the paragraph vector is learned by holding the softmax parameters and gradient descending on the new vector entry.

\begin{equation}
\label{softmax}
p(w_{i}|w_{i-c},...,w_{i+c})=\frac{e^{y_{w_{i}}}}{\sum_{j\in (1,...,T)} e^{y_{wj}}}
\end{equation}

\subsection{Contextual Features Enrichment}
Simply using documents that are in a large scale into the models is hardly good enough for learning a good representation of the documents \cite{ahn2016neural}.  \emph{doc2vec} model can learn very good representations of the documents semantically. However, in some domains like job boards, many documents can share major overlapping content such as company or benefit descriptors, while the important context including the distinguished information such as requirements or qualifications are overshadowed. For example, in job boards domain 90\% of a job description may be dedicated to describe the company and its values and culture, while only 10\% describes the job requirements. In such scenario, \emph{doc2vec} will generate almost same representation for two jobs posted by the same company, however they are totally different by the job requirements. To overcome this problem, we enrich each document with contextual features including the document classification, and location. Additionally, in our domain since not all the parts of a document (job posting) is equally important to measure similarity, we utilized our in-house NLP document parser to extract the important content such as job requirements and skills. These extracted information is injected into the original document $N$ times (we choose $N$ to be 3 empirically) to guide \emph{doc2vec} to a better representation by improving the content distribution.

\section{Experiment and Results}
\label{exp}
To validate and evaluate the proposed technique, we applied our approach on top of Careerbuilder's CF-based recommendation engine. CareerBuilder operates one of the largest job boards in the world and has an extensive and growing global presence, with millions of job postings, more than 60 million actively-searchable resumes, over one billion searchable documents, and more than a million searches per hour. In their recommendation engine, Careerbuilder strives to recommend the right job to the right person at the right time. However, the cold-start problem is very serious at Careerbuilder given that every day thousands of new jobs are posted and it is important for the employers who have posted those jobs to start receiving applications from job seekers within a short period of time. Recommendation is a major channel of job exposure, therefore new jobs will lose a chance to be exposed through this important channel due to the cold-start problem.
In this section we will discuss the different experiments we run to evaluate solving the cold-start problem using our technique. All experiments are done in a machine with Intel(R) Xeon (R) E5-2667 series CPUs (32 cores in total) and 264GB memory. The runtime of each model is also evaluated since our application scenario has a requirement of running the whole workflow on a large dataset (more than one million documents) on a daily basis.

\begin{figure}[h]
	\centering	
	\includegraphics[width=3.4in]{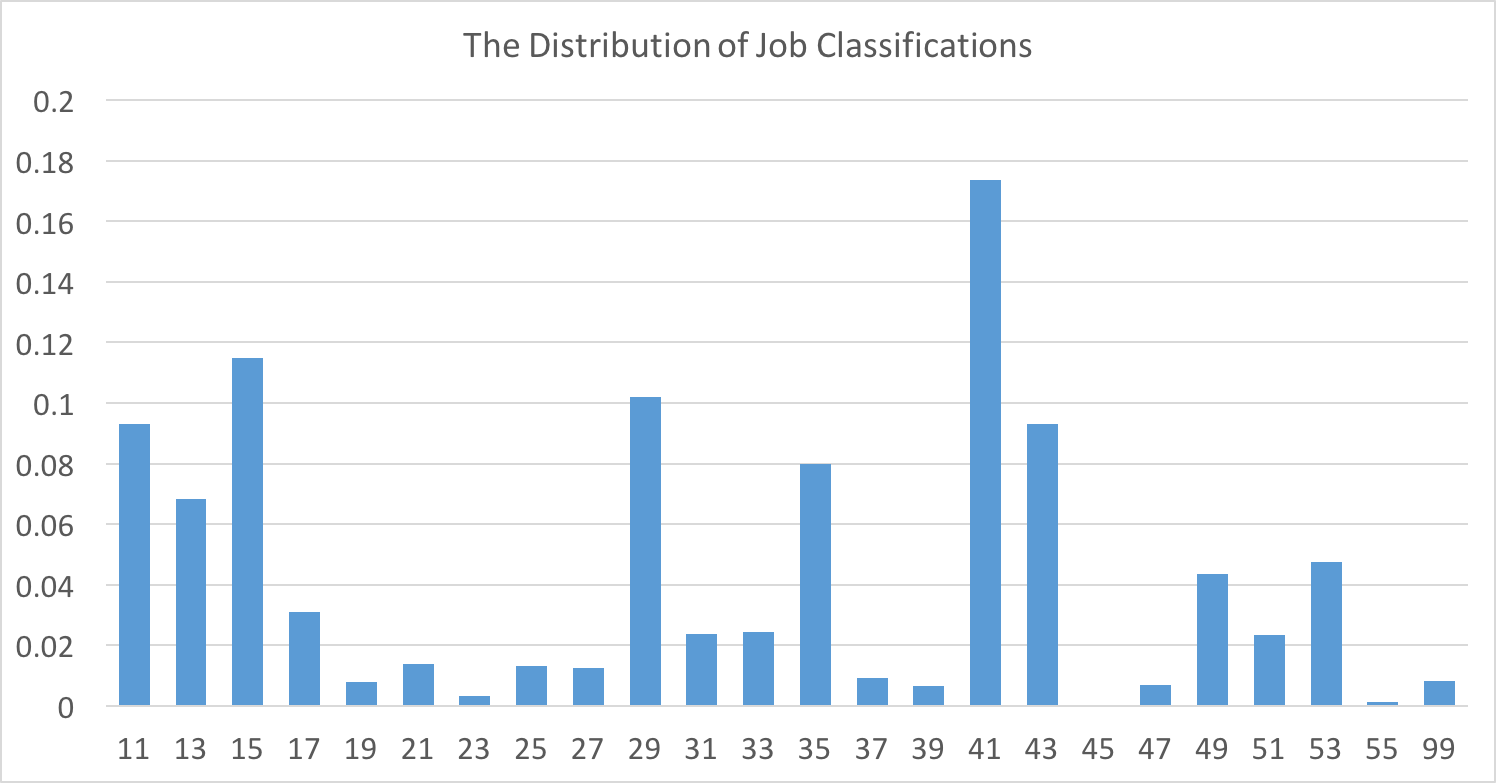}	
	\caption{Distribution of 24 General Job Classifications: x-axis denotes the job category code and y-axis denotes the proportion of jobs under each category.}
	\label{distribute}
\end{figure}

\begin{table*}[th]
	\centering
	\caption{Recall for Top 10-50 Most Similar Jobs.}
	\label{result}
	\begin{tabular}{ccccc}
		\hline
		Model & Top 10 & Top 20 & Top 30 & Top 50 \\ \hline\hline
		\emph{tf-idf} & 30.4\% & 37.3\% & 39.9\% & 44.8\% \\ \hline
		\emph{LDA} & 18.2\% & 21.4\% & 26.8\%  &  30.2\%  \\ \hline
		\emph{doc2vec} & 88.9\% & 95.0\% & 96.3\%  & 97.7\%  \\ \hline
		\emph{tf-idf} with contextual features & 32.1\% & 41.2\% & 46.4\%  &  49.5\%  \\ \hline
		\emph{LDA} with contextual features & 19.9\% & 25.3\% & 30.1\%  & 33.3\%   \\ \hline
	\end{tabular}
\end{table*}
\subsection{Job-to-Job (J2J) Matcher}
All experiments used a sample of 1,147,725 classified jobs from our dataset~\cite{korayem2015query}. The distribution of job classification is shown in Figure \ref{distribute}. For experiments we use Gensim \cite{rehurek_lrec} to generate \emph{tf-idf} model and train \emph{LDA} and \emph{doc2vec} models. For the \emph{doc2vec} model, we train our model with one iteration only due to speed limitation in order to run the whole process on a daily basis, and chose 100 dimensions for the embedded vectors. The reason is that, according to our experiments adding more dimensions for the vectors barely improved the performance if any, but increased the runtime notably. We choose the number of job categories which have at least 100 jobs included in our testing dataset to be the number of topics used for generating \emph{LDA} model, so that one topic is expected to indicate one job category (java developer, registered nurse, etc). Eventually we have 805 topics in total. We use the fine-grained categories instead of the general ones since the number is small and can hardly learn a very good representation of the documents. We chose cosine similarity as the similarity metric to evaluate similarity between documents with threshold 0.5 to indicate if two documents are similar or not.

In this section we test the embedding algorithms for learning document level similarity, among which the \emph{tf-idf} and \emph{LDA} models are set as baselines. We evaluated the performance of each model both with and without contextual features enrichment. Since the contextual feature enrichment is proposed to improve the performance for certain specific circumstances, we don't add it in this experiments section. Our objective is to evaluate the job-to-job matching performance for our model rather than to evaluate the results for the whole recommendations, i.e., to evaluate the similarity we learned from different models.

\begin{table}[h]
	\centering
	\caption{Precision for Top 10 Most Similar Jobs.}
	\label{prec}
	\begin{tabular}{cc}
		\hline
		Model & Precision\\ \hline\hline
		\textit{tf-idf} & 29.2\% \\ \hline
		\textit{LDA} & 17.1\% \\ \hline
		\textit{doc2vec} & 82.9\% \\ \hline
		\textit{tf-idf} with contextual features & 30.3\% \\ \hline
		\textit{LDA} with contextual features & 18.4\% \\ \hline
		\textbf{\textit{doc2vec} with contextual features} & \textbf{91.8\%} \\ \hline
	\end{tabular}
\end{table}

We started our evaluation by selecting 100 randomly selected jobs as test samples and generated the top 10 most similar jobs for each test sample based on the \emph{doc2vec} model with contextual features to achieve the best performance and evaluated the result manually as good/bad matches. The good matches are then considered as ground truth for evaluating other models.
The overall precision for this set-up is 91.8\%. To evaluate the other models, we use the similar jobs which have been labeled as good matches as ground truth and calculate the recall value (namely, among all the correct matches, how many of them have been returned as matches by the other models) from top 10 to 50 most similar jobs generated by the other models, as well as the precision based on the top 10 most similar jobs returned from each model (namely, among the top 10 most similar jobs returned from the other models, how many of them are labeled as correct match in our evaluation). The results are shown in Table \ref{result} and Table \ref{prec}.

\begin{table}[h]
	\centering
	\caption{Running Time for Different Model.}
	\label{time}
	\begin{tabular}{cc}
		\hline
		Model & Running Time (Minutes) \\ \hline \hline
		\textit{tf-idf} & 22 \\ \hline
		\textit{LDA} & 992 \\ \hline
		\textit{doc2vec}  & 75 \\ \hline
		\textit{tf-idf} with contextual features & 30 \\ \hline
		\textit{LDA} with contextual features & 1069 \\ \hline
		\textit{doc2vec}  with contextual features & 89 \\ \hline
	\end{tabular}
\end{table}

 \begin{table*}[th]
 	\centering
 	\caption{Case Study: Source Job (left), Poor Pairing (middle) and Good Pairing (right).}
 	\label{case}
 	\begin{tabular}{|C|D|c|c|}
 		\hline
 		\textbf{Document Title:} & HVAC Technician
 		
 		XX Resort \& Club & Banquet Houseperson - PM & HVAC Mechanic \\ \hline
 		\textbf{\begin{tabular}[c]{@{}c@{}}Same Company \\ Values Context\end{tabular}} & \multicolumn{3}{W|}{\begin{tabular}[c]{@{}W@{}} \\[-10pt]Since being founded in 1919, Company XX has been a leader in the hospitality industry. Today, Company XX remains a beacon of innovation, quality, and success. This continued leadership is the result of our Team Members staying true to our Vision, Mission, and Values. Specifically, we look for demonstration of these Values: H Hospitality - We're passionate about delivering exceptional guest experiences... In addition, we look for the demonstration of the following key attributes in our Team Members: Living the Values Quality Productivity Dependability Customer Focus Teamwork Adaptability.\end{tabular}} \\[-10pt]\\ \hline
 		
 		\textbf{\begin{tabular}[c]{@{}c@{}}Same Company\\ Overview Context\end{tabular}} & \multicolumn{3}{W|}{\begin{tabular}[c]{@{}W@{}} \\[-10pt] What will it be like to work for this Company XX Brand? What began with the world's most iconic hotel is now the world's most iconic portfolio of hotels. In exceptional destinations around the globe, XX Hotels \& Resorts reflect the culture and history of their extraordinary locations, as well as the rich legacy of Company XX... If you understand the value of providing guests with an exceptional environment and personalized attention, you may be just the person we are looking for to work as a Team Member with XX Hotels \& Resorts.\end{tabular}} \\[-10pt]\\ \hline
 		
 		\textbf{\begin{tabular}[c]{@{}c@{}}Similar Benefits \\ Context\end{tabular}} & \multicolumn{3}{W|}{\begin{tabular}[c]{@{}W@{}} \\[-10pt]What benefits will I receive? Your benefits will include a competitive starting salary and, depending upon eligibility, a vacation or Paid Time Off (PTO) benefit. You will instantly have access to our unique benefits such as the Team Member and Family Travel Program, which provides reduced hotel room rates at many of our hotels for you and your family, plus discounts on products and services offered by Company XX Worldwide and its partners. After 90 days you may enroll in Company XX Worldwide Health \& Welfare benefit plans, depending on eligibility. Company XX Worldwide also offers eligible team members a 401K Savings Plan, as well as Employee Assistance and Educational Assistance Programs. We look forward to reviewing with you the specific benefits you would receive as a Company XX Worldwide Team Member...\end{tabular}} \\[-10pt] \\ \hline
 		
 		\textbf{Differences:} & \begin{tabular}[c]{@{}T@{}}
 			\\[-10pt]
 			The XX Resort \& Club is seeking \textbf{an HVAC Technician who will be responsible for maintaining the physical functionality and safety of the hotels heating, ventilation and air conditioning (HVAC) equipment and machinery continuing effort to deliver outstanding guest service and financial profitability.} \\ What will I be doing? \\	\textbf{As an HVAC Mechanic, you would be responsible for maintaining the physical functionality and safety of the hotels heating, ventilation and air conditioning (HVAC) equipment and machinery in the hotels continuing effort to deliver outstanding guest service and financial profitability. Specifically...} \\[-10pt]\\ 
 		\end{tabular} &
 		
 		\begin{tabular}[c]{@{}L@{}} As a Banquet Set-Up Attendant, you would be responsible setting and cleaning banquet facilities for functions in the hotel's continuing effort to deliver outstanding guest service and financial profitability. Specifically, you would be responsible for performing the following tasks to the highest standards: Set tables and chairs to meet function specifications. Clean meeting space including, but not limited to, vacuuming, sweeping, mopping, polishing, wiping areas and washing walls before and after events...\end{tabular} &
 		
 		\begin{tabular}[c]{@{}T@{}}\textbf{An HVAC Mechanic with Company XX Hotels and Resorts is responsible for maintaining the physical functionality and safety of the hotels heating, ventilation and air conditioning (HVAC) equipment and machinery continuing effort to deliver outstanding guest service and financial profitability.} \\ \textbf{As an HVAC Mechanic, you would be responsible for maintaining the physical functionality and safety of the hotels heating, ventilation and air conditioning (HVAC) equipment and machinery in the hotels continuing effort to deliver outstanding guest service and financial profitability. Specifically...} 
 		\end{tabular} \\ \hline
 	\end{tabular}
 \end{table*}

According to the results shown in Table \ref{prec}, our proposed \emph{doc2vec} model with contextual features yields the best performance for job-to-job matching task. On the other hand, the \emph{LDA} model obtains the lowest. The \emph{LDA} model learns a good representation of the documents for classifying them under different topics, however for job recommendation and job-to-job pairing tasks it does not perform good enough. Previous works have shown that \emph{tf-idf} model itself is capable of learning a good document embedding for documents which are long enough, and according to our results the \emph{tf-idf} model outperforms the \emph{LDA} model by about 10\%. Since the documents in our dataset are mainly job descriptions with contextual features including job requirements, job title, job classifications, etc., which are not expected to be long enough for \emph{tf-idf} to learn a very good representation. The \emph{doc2vec} outperforms the \emph{tf-idf} and \emph{LDA} models significantly and achieves the highest precision of 91.8\%. By including the contextual features we achieve a 9\% gain in the precision of the top-10 most similar jobs generated by the \emph{doc2vec} model. For \emph{LDA} and \emph{tf-idf} the improvements are not comparable because the two baseline models are not capable of learning a good embedding for the documents and adding more information does not help. However, on the other hand the \emph{doc2vec} model can learn a decent document embedding semantically and by adding the extra information we are able to obtain that significant improvement on performance. Table \ref{result} shows the recall value of different models from top 10 to 50 most similar jobs. Since the manually labeled results based on \emph{doc2vec} with contextual features are established as ground truth, we did not calculate the recall value for it. Based on our results the \emph{doc2vec} model without contextual features still performs better than the other models, followed by the \emph{tf-idf} model and then the \emph{LDA} model. The \emph{doc2vec} model without contextual features is able to return about 90\% of the correct job matches for top 10 most similar jobs and almost all of the correct matches with the top 50, while the \emph{tf-idf} model reaches at about 40\% and the \emph{LDA} model reaches at about 30\%. By adding the context we still obtain an improvement on the recall values which is consistent with the results on precision.

As introduced in Section \ref{method}, we have thousands of new jobs added to the system on a daily basis and about the same number of jobs expiring. Therefore it is best to train our model and generate similar jobs for those jobs which suffer from the cold-start problem on a daily basis. This would require the runtime for the whole process including training and inferring to be in an hourly-scale. The runtime for each model is shown in Table \ref{time}. We can conclude that by using multi-process computing (all models are using 24 cores for multi-process computing) we can run all the models on a daily basis except the \emph{LDA} model. contextual features adds more text contents and more computation in the process, but with multi-process computing the extra runtime is reasonable compared with the significant improvement on the performance. The runtime for \emph{tf-idf} model is the shortest, while \emph{doc2vec} model with and without contextual features is still comparable which can be completed within two hours. For a daily based workflow it is acceptable and we choose to run it at midnight every day for our job recommendation systems.

\begin{table*}[!] \centering \caption{Selected Parts of a
		User-to-Job Matching Example.} \label{case-r2j}
	\begin{tabular}{|c|Y|c|J|} \hline
		\multicolumn{2}{|c|}{\textbf{Resume}} &
		\multicolumn{2}{c|}{\textbf{Job}}                  \\ \hline
		Education & \begin{tabular}[c]{@{}Y@{}} Master of Business
			Administration Degree in Accounting. \end{tabular} & Title &
		\begin{tabular}[c]{@{}K@{}} Senior Accountant - General Ledger(GL)
		\end{tabular} \\ \hline
		
		Certificate & \begin{tabular}[c]{@{}Y@{}} Candidate for
			CPA (passed BEC and Regulations). \end{tabular} &
		\multirow{2}*{Requirements} &
		\multirow{2}{*}{{\begin{tabular}[c]{@{}K@{}} \begin{itemize}
						\item Must be a degreed accountant. \item CPA and/or Master
						of Science in Accounting. \item Minimum of 2-3 years general
						accounting experience in a corporate division or Big Four
						environment is required. \item Must demonstrate proficiency
						in general ledger accounting including preparing, reading,
						interpreting, and analyzing financial statements. \item...
					\end{itemize} \end{tabular}}}
					
					\\ \cline{1-2} Skills& \begin{tabular}[c]{@{}Y@{}}
						\begin{itemize} \item Expert in handling full accounting
							cycle operations with hands on experience in receivables,
							payable, bank reconciliation, tax planning and filing, etc.
							\item Successful project manager with expertise to
							effectively budget and forecast resource needs, plan and
							schedule time and resources for meeting expectations. \item
							... \end{itemize} \end{tabular} &  & \\ \hline
					
					Work Experience& \begin{tabular}[c]{@{}Y@{}}
						\begin{itemize} \item Senior Accountant 12/2015-Present \item
							Senior Accountant Consultant 11/2015-2/2016 \item Accounting
							Consultant 7/2015-11/2015 \item Accounting Manager
							5/2014-3/2015 \item Senior Accountant 7/2013-5/2014 \item ...
						\end{itemize} \end{tabular}
						
						& Responsibilities& \begin{tabular}[c]{@{}K@{}}
							\begin{itemize} \item Maintain project commitments ensuring
								proper accounting treatment for all projects. \item Compile
								and analyze financial information to interpret and
								communicate current and projected company financial results
								to management. \item Prepare and examine financial statements
								and footnotes as assigned for completeness, accuracy and
								conformance with relevant accounting standards and management
								reporting requirements. \item... \end{itemize} \end{tabular}
						\\ \hline \end{tabular} \end{table*}

\subsection{Case Study}
The job-to-job matcher did admirably matching documents which are similar to one another, however, we encountered challenges with documents comprised of approximately 70\% or higher of exact same content due to company descriptors like
background, benefits and values. As the example shown in Table \ref{case}, compared with the source job on the left column, HVAC Technician was matched to Banquet Houseperson in the middle with completely different skill sets and requirements. This is due to the document being so similar as stated previously. In this example 25\% of the document is actually different which still allows poor recommendations which are highly similar to prevail and be recommended. To counter, we enrich contextual features which helped alleviate and push down similarity scores for these edge cased documents. Referring to the right column, HVAC Mechanic was returned instead when we emphasized on promoting the differences within the document. In our domain, the contextual features proved successful in filtering out highly similar scoring documents with different roles and functions.


\section{Conclusions and Future Work}
\label{discussion}
Collaborative Filtering is widely used in recommendation systems for real world applications, however, it suffers from the cold-start problem where new entities cannot be recommended or receive recommendations including the users and items. To tackle the cold-start problem we build an item-to-item deep learning matcher based on the document similarities learned from the state-of-the-art document embedding model \emph{doc2vec}. The performance of document level similarities learned by the \emph{doc2vec} model with contextual features outperforms the baseline models significantly and can run on a daily basis which fits the requirement of our practical application. Our approach can be integrated with any existing CF-based recommendation engine with no need to modify the CF core. To proof the efficiency, scalability, and accuracy of the proposed technique we apply it on top of Careerbuilder's CF-based recommendation engine which is used to recommend jobs to job seekers. After testing this model on more than 1 million documents we prove its efficiency in resolving the cold-start problem in large scale while maintaining high level of accuracy. We are working on a multimodal document embedding model for learning user-to-user and user-to-job similarity whose initial results are very promising to solve the cold-start problem for user-based CF as shown in Table \ref{case-r2j}.

%
%

%
%

\bibliographystyle{ieeetr}
\bibliography{ref}
\end{document}